%% file: main.tex
\documentclass[runningheads]{llncs}
\input{latex_files/packages}
\input{latex_files/macro}

\usepackage[T1]{fontenc}
\usepackage{graphicx}

\begin{document}
\title{Enhancing Privacy in Federated Learning: Secure Aggregation for Real-World Healthcare Applications} 
\titlerunning{Secure Aggregation for Real-World Healthcare Applications}
%
\author{Riccardo Taiello\inst{1,2,3}\orcidID{0000-0002-9890-9639} \and
Sergen Cansiz\inst{1} \and Marc Vesin\inst{1} \and Francesco Cremonesi\inst{1} \and Lucia Innocenti\inst{1}
\and Melek Önen\inst{2}\orcidID{0000-0003-0269-9495} \and Marco Lorenzi \inst{1,3} \orcidID{0000-0003-0521-2881}}
\authorrunning{Taiello et al.}
\institute{Epione Research Project, Inria, Sophia Antipolis, France \and
EURECOM, Sophia Antipolis, France\\
\and
Université Côte d’Azur, Nice, France
\email{\{riccardo.taiello,marco.lorenzi\}@inria.fr} \\ \email{melek.onen@eurecom.fr}}
\maketitle              
\begin{abstract}
Deploying federated learning (FL) in real-world scenarios, particularly in healthcare, poses challenges in communication and security. In particular, with respect to the federated aggregation procedure, researchers have been focusing on the study of secure aggregation (SA) schemes to provide privacy guarantees over the model's parameters transmitted by the clients. Nevertheless, the practical availability of SA in currently available FL frameworks is currently limited, due to computational and communication bottlenecks. To fill this gap, this study explores the implementation of SA within the open-source Fed-BioMed framework. 
We implement and compare two SA protocols, Joye-Libert (JL) and Low Overhead Masking (LOM), by providing extensive benchmarks in a panel of healthcare data analysis problems. Our theoretical and experimental evaluations on four datasets demonstrate that SA protocols effectively protect privacy while maintaining task accuracy. Computational overhead during training is less than 1\% on a CPU and less than 50\% on a GPU for large models, with protection phases tacking less than 10 seconds. Incorporating SA into Fed-BioMed impacts task accuracy by no more than 2\% compared to non-SA scenarios.
Overall this study demonstrates the feasibility of SA in real-world healthcare applications and contributes in reducing the gap towards the adoption of privacy-preserving technologies in sensitive applications.

\end{abstract}
\section{Introduction}
\input{latex_files/introduction}
\section{Background}
\input{latex_files/background}

\section{Methods}
\input{latex_files/method}
\section{Evaluation}
\input{latex_files/evaluation}
\section{Conclusion and Future Works}
We have demonstrated that SA can be effectively implemented within the Fed-BioMed framework to enhance privacy in federated learning. Our evaluations using four medical datasets show that both Joye-Libert and Low Overhead Masking protocols protect privacy while maintaining task accuracy. The computational overhead is minimal, making SA a viable option for real-world deployments. 
As part of future work, we plan to replace MP-SPDZ with a direct implementation of additive secret sharing within Fed-BioMed. We also aim to replace JL with a quantum-resistant SA \cite{brakerski2011fully} using the \textsc{SHELL} C++ library \footnote{{\textcolor{blue}{SHELL library: \url{https://github.com/google/shell-encryption/}}}}.
\newpage
\subsubsection*{Acknowledgements}
This work has been supported by the French government, through the 3IA Côte
d’Azur Investments in the Future project managed by the National Research Agency (ANR) with the reference number ANR-19-P3IA0002, by the TRAIN project ANR-22-FAI1-0003-02, and by the ANR JCJC project Fed-BioMed 19-CE45-0006-01.
%
%
%
\bibliographystyle{splncs04}
\bibliography{bibliography}
\appendix
\input{latex_files/appendix}

\end{document}

%% file: latex_files/packages.tex
\usepackage{amsfonts} 

\usepackage{url}
\usepackage[hidelinks]{hyperref}
\usepackage{adjustbox}
\usepackage{nicematrix}

\usepackage{graphicx}

\usepackage{booktabs}
\usepackage{multirow}
\usepackage{colortbl}

\usepackage{enumitem}
\usepackage{xspace}
\usepackage{caption}
\usepackage{comment}

\usepackage{bm}
\usepackage{pifont}

\usepackage{graphicx} 
\usepackage{subcaption} 

%% file: latex_files/macro.tex
\newcommand{\tbf}[1]{\text{\textbf{#1}}}
\newcommand{\users}{{\mathcal{U}}}
\newcommand{\selected}{\mathcal{U}^{(\tau)}}

\newcommand{\uuu}{\forall u \in \users}

\newcommand{\rgets}{\xleftarrow{R}}

\definecolor{Gray}{gray}{0.92}

\newcommand{\vinputsync}{\vec{x}_{u,\tau}}

\newcommand{\vencsync}{\vec{y}_{u,\tau}}

\newcommand{\vaggsync}{\vec{x}_{\tau + 1}}

%% file: latex_files/introduction.tex
Federated Learning (FL) is a distributed machine learning paradigm that enables multiple clients to collaboratively train a global model without sharing their local datasets. While researchers have largely focused in developing FL theories and methods in a variety of applications, the deployment of FL in real-world scenarios is still challenging, particularly in terms of communication protocols, security, and customization bottlenecks.

A critical requirement for real-world applications of FL concerns the protection of the model's parameters shared by the clients during model aggregation. To this end, privacy-preserving methodologies such as Secure Aggregation (SA) \cite{mansouri2023sok} are currently under study, to guarantee that aggregated data shared among participants do not reveal individual contributions. Contrarily to other privacy-enhancing technologies like Differential Privacy (DP) \cite{dwork2006differential}, the privacy guarantees of SA rely on the security proofs of established cryptographic primitives \cite{JL2013,kursawe2011privacy}. 

On the practical side, while DP requires only minor adjustments to the federated aggregation process through the injection of noise to the model's parameters, implementing SA in production is more complex as it requires changes to the standard operational flow of the FL framework by incorporating new communication phases. As a result, the adoption of SA in currently available FL software frameworks is lagging behind. Existing SA solutions primarily target settings with a large number of clients, where hardware limitations can lead to protocol execution failures. Some preliminary solutions have been proposed in the framework \textsc{Flower} \cite{beutel2020flower}, which however introduce a non-negligible overhead. 
The approach provided by \textsc{NVFlare} is simpler but suffers from a weak security model \cite{roth2022nvidia}. Finally SA in \textsc{OpenFL} \cite{reina2021openfl} requires dedicated hardware solutions. Overall, the applications of these SA protocols in the cross-silo healthcare setting is suboptimal, due to the limited number of clients, and their general availability as compared to the cross-device setting.

To address these limitations, in this work we explore the implementation of SA schemes optimally customized for cross-silo healthcare applications. In particular, we study the two suitable categories of SA based on masking and additively homomorphic encryption \cite{mansouri2023sok}.  We identify respectively \textsc{Low Overhead Masking} \cite{kursawe2011privacy} and \textsc{Joye-Libert} \cite{JL2013} as the most relevant solutions for our application. These protocols are designed to protect individual updates from being exposed during the aggregation process. 

This work is based on theoretical and experimental evaluation of these SA protocols within the Fed-BioMed framework \cite{cremonesi2023fed}. 
In particular, we conducted a comprehensive comparison on four distinct medical datasets including medical images and tabular data: Fed-IXI \cite{ogier2022flamby}, Fed-Heart \cite{ogier2022flamby}, REPLACE-BG \cite{aleppo2017replace}, and FedProstate \cite{innocenti2023benchmarking}. We measured the computational resources required for training, encryption, and overall execution time. When training was performed on a CPU, we achieved a total computation overhead of less than 1\%, while on a GPU, for larger machine learning models ($>5M$ parameters), the overhead was less than 50\%, with a protection phase that took less than 10 seconds. Furthermore, we analyzed the impact of SA on task accuracy, demonstrating that incorporating SA into Fed-BioMed affects accuracy by no more than 2\% compared to non-SA scenarios. 
Overall this study demonstrates the feasibility of SA in real-world healthcare applications and contributes in reducing the gap towards the adoption of privacy-preserving technologies in sensitive applications.

%% file: latex_files/background.tex
\paragraph{Federated Learning.}
As introduced by McMahan et al. \cite{mcmahan17fl}, FL consists of a distributed machine learning paradigm where a group of clients, denoted as $\users$, collaboratively trains a global model with parameters $\vec{\theta} \in \mathbb{R}^d$, under the guidance of a FL server. One of the first and popular methods used to train a FL model is the FedAvg scheme \cite{mcmahan17fl}. With FedAvg, at each FL round denoted by $\tau$, each client $u \in \users$ 
trains the  model $\vec{\theta}_{u,\tau}$ on the private local data $\mathcal{D}_u$, for example through Stochastic Gradient Descent (SGD) \cite{sgd-2004}. Upon completion of the local training, each client forwards its updated model $\vec{\theta}_{u,\tau}$ to the server and the local dataset size $w_u = |\mathcal{D}_u|$.
When the server receives the updated models from all participating clients, it proceeds to the weighted aggregation step:
  \begin{equation*}
\vec{\theta}_{\tau+1} \gets \frac{\sum_{\uuu}{w_u \vec{\theta}_{u,\tau}}}{\sum_{\uuu} w_u} .
\end{equation*}

This iterative process continues until the global model $\vec{\theta}$ reaches some desired level of accuracy. 
The presence of a large number of FL clients significantly impacts the communication overhead. To mitigate this, instead of involving all clients in the training, at each FL round, the server selects a subset of clients (\textit{client selection} \cite{mcmahan17fl}), denoted as $\selected \subseteq \users$, with $|\selected| = n$, and collects their parameters only for aggregation.
\paragraph{Secure Aggregation.}
SA \cite{mansouri2023sok} typically involves multiple \emph{users} and a single \emph{aggregator}. Each user possesses a private input, and the role of the aggregator is to calculate the sum of these inputs. A property of SA is that the aggregator learns nothing more than the aggregated sum, thereby preserving the privacy of individual user inputs.

SA has found significant applications in Federated Learning (FL), where it is used to securely aggregate the updated model parameters received from FL clients (aligned with the \emph{user's} concept in SA) during each FL round, by instantiating an FL server (\emph{aggregator} in the context of SA). The adoption of SA is motivated by the potential threats posed by adversaries having access to the client's updated model $\vec{\theta}_{u,\tau}$ which may infer information about its private dataset $\mathcal{D}_u$ \cite{milad19,InferenceAttack17}. Hence, the local models should remain confidential even against the FL server. 
SA in FL was first developed by Bonawitz et al. \cite{googleccs17}. The  protocol considered in that study faced two different challenges:

$\bullet$ \textit{Threat models} defining the potential risks and behaviors that the security protocol is designed to protect against. The primary threat scenarios in SA include the honest-but-curious model where parties (server and clients) follow the protocol without tampering with the data but may attempt to infer additional information.

$\bullet$ \textit{Client dropouts}, caused by factors such as connectivity issues or voluntary withdrawal, are common in real-world federated learning environments. Dropouts can significantly impact the computation and number of communication rounds of the SA protocol, as they often require the participation of all selected clients within a training round. With communication rounds, we refer to the number of interactions required between the clients and the server to complete a particular phase of the protocol.

\section{Related Works}

In real-world deployments, only a few FL frameworks implement some form of SA: \textsc{OpenFL}\cite{reina2021openfl}, \textsc{NVFlare}\cite{roth2022nvidia}, and \textsc{Flower} \cite{beutel2020flower}.

\textsc{Flower} implements \textsc{SecAgg+} \cite{bell2020secure}, a masking-based protocol that ensures security in the honest-but-curious model. This protocol requires four communication rounds and uses Shamir's Secret Sharing to recover missing masks in case of client dropout, ensuring the server can complete the aggregation. Compared to the SA schemes here introduced in Fed-BioMed, Flower's approach is more costly in terms of communications, albeit accommodating for client dropout.

\textsc{NVFlare} introduces an SA method that leverages the CKKS asymmetric homomorphic encryption scheme \cite{cheon2017homomorphic}. This threat model is considered weaker than typical state-of-the-art protocols because it requires clients to share a common secret key and assumes clients are honest. Clients protect their inputs using a public key, while the server, operating under the honest-but-curious model, aggregates these inputs and returns the aggregate to each client for decryption using the same secret key. This approach requires one communication round and allows client dropout. 

\textsc{OpenFL}’s use of Trusted Execution Environments (TEEs) represents a further step in sandboxing and securing local computations, but requires specific hardware which may not be available in typical FL studies involving hospitals.

%% file: latex_files/method.tex
\begin{figure}
    \centering
    \includegraphics[width=1.0\textwidth]{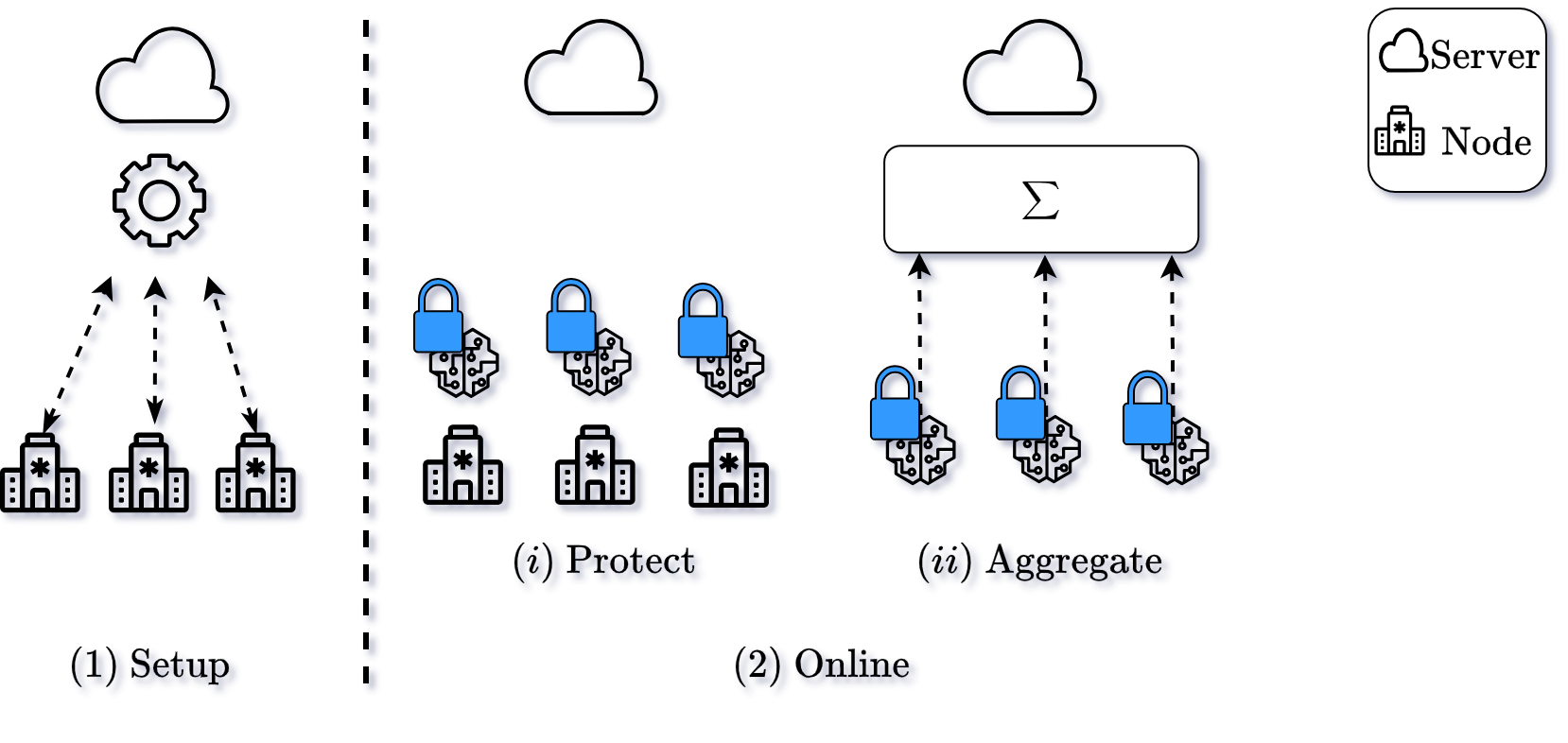}
    \caption{Overview of Secure Aggregation phases.}
    \label{fig:sa}
\end{figure}
In this section, we detail the implementation in Fed-BioMed of the two SA protocols, \textsc{Joye-Libert} (JL) and \textsc{Low Overhead Masking} (LOM). From this point on, we adopt the terminology of Fed-BioMed, where a client is referred to as a node. 

A general overview of SA is depicted in Figure \ref{fig:sa}, and a more detailed scheme is provided in Supplementary Figure \ref{fig:sa_detailed}. An SA protocol comprises two phases: \textbf{setup} and \textbf{online}. The setup phase, illustrated in Figure \ref{fig:sa}.1, is executed among all participating nodes in \(\users\) before the FL training. This step ensures that all parties have the appropriate cryptographic material necessary to run the specific SA protocol.

The online phase, Figure \ref{fig:sa}.2 is repeated during each FL round \(\tau\) and consists of two steps: $(i)$ \textit{protect} and $(ii)$ \textit{aggregate}. In the \textit{protection} step, each node protects its private local model using specific SA primitives and then sends the protected model to the server. In the \textit{aggregation} step, the server receives the protected local models, computes the aggregate, and then decrypts it. To ensure the correct functioning of the cryptographic primitives, the locally-trained model vector of each node must be quantized beforehand.

\paragraph{Prerequistes.}
To perform FedAvg with SA, we first convert the node's local parameters \(\Vec{\theta}_{u, \tau} \in \mathbb{R}^d\) into integers \(\mathbb{Z}^d_{2^L}\), where \(L\) represents the maximum number of bits of the plaintext. This conversion is achieved by applying uniform quantization, defined as:
$Q(\Vec{\theta}_{u,\tau}) = \left\lfloor \frac{2^L \cdot (\Vec{\theta}_{u,\tau} - \theta_{\min})}{(\theta_{\max} - \theta_{\min})} \right\rceil.$ Here, \(\lfloor \cdot \rceil\) denotes the standard rounding function. To ensure that real values are within a desired range, we apply a clipping function, \(\text{clip}(x, \theta_{\min}, \theta_{\max})\) = \(\min(\max(x, \theta_{\min}), \theta_{\max})\), where \(\theta_{\min}\) and \(\theta_{\max}\) are the lower and upper bounds, respectively.

To apply weighted averaging over the integers, we assume that \(w_u \in \mathbb{Z}_{2^{W_u}}\), where \(W_u\) is the number of bits to represent the node's dataset size, and we define \(W = \max(\{W_u\}_{\uuu})\).

The weighted local model is computed as \(\vinputsync = Q(\Vec{\theta}_{u,\tau}) \cdot w_u\), resulting in \(\vinputsync \in \mathbb{Z}^d_{2^{L + W}}\). To avoid overflow, we define \(M = L + W + \log_2(n)\) as the maximum number of bits for sum computation. The aggregate \(\vaggsync = (\sum_{u \in \selected} \vinputsync) \in \mathbb{Z}^d_{2^M}\) is then divided by \(s = \sum_{u \in \selected} w_u\) and dequantized using the following formula: $
\Vec{\theta}_{\tau+1} = Q^{-1}(\vaggsync) = \vaggsync \cdot \frac{(\theta_{\max} - \theta_{\min})}{2^L} + \theta_{\min}.$

In this context, we assume that quantization has been performed and omit the details of the dequantization process in the protocol explanation.

\subsection*{Joye-Libert}

In Supplementary Figure \ref{fig:protocol_v1}, we illustrate the Joye-Libert (JL) implementation in Fed-BioMed. During the \textbf{setup} phase, the participating nodes \(\users\) generate their private keys \(sk_u\), and the server creates its server key \(sk_0\) — which is the sum of the node keys — using Shamir Secret Sharing (SS) \cite{shamir79}. 

During the \textbf{online} phase, the protection and aggregation are applied as described in JL (Section 4 \cite{JL2013}). In the protection step, each node uses a private secret key \(sk_u\) at FL round \(\tau\) with a one-time mask derived from \(sk_u\) and \(\tau\), to obtain a protected local model through modular exponentiation over a large moudulus $N$. Using the server key \(sk_0\), the server can recover the aggregate of the nodes' private local models in clear.

Our JL implementation works with vectors; the protection and aggregation algorithms are applied element-wise. We use the element's index \(i\) to generate a unique FL round (need to guarantee a one-time mask) for each element in the vector. For instance, to protect \(\vinputsync\), we execute protect and aggregate over the FL round \(\tau || i\) and input \(\vinputsync[i]\), where \(\vinputsync[i]\) represents the \(i\)-th element of the vector \(\vinputsync\).

The computation and communication of the protected local model is optimized by using vector encoding \cite{mansouri2022learning}.

\paragraph{Software details.}
SS is integrated into Fed-BioMed using MP-SPDZ library \cite{keller2020mp}. The modulus $N$ is provided by the server, and the modular operations are performed using the \textsc{gmpy2}\footnote{{\textcolor{blue}{\textsc{gmpy2}: \url{https://gmpy2.readthedocs.io/}}}} Python library.

\subsection*{Low Overhead Masking}

The second implementation, Low Overhead Masking (LOM) \cite{kursawe2011privacy}, which supports client selection, is depicted in Supplementary Figure \ref{fig:protocol_v2}. 
During the \textbf{setup} phase, all participating nodes \(\users\) establish a pairwise secret \(s_{u,v}\), such that \(s_{u,v} = s_{v,u}\), with all nodes through the Diffie-Hellman Key Agreement (KA) \cite{diffie2022new}, which will be used in the protect step.

In the \textbf{online} phase, during the \textit{protection} step, a selected node \(u \in \selected\) runs the protect algorithm (Section 3.4 \cite{kursawe2011privacy}). This algorithm protects the local model with a one-time mask derived through a Pseudo-Random Function (PRF) which uses the pairwise secret with the selected nodes \(\selected\) and the current FL round \(\tau\), and sends the protected local model to the server. The server then sums the protected local models and collects the final aggregate \(\vaggsync\).

\paragraph{Software details.}
Diffie-Hellman KA and PRF are implemented in the \textsc{cryptography}\footnote{\textcolor{blue}{ \textsc{cryptography}:  \url{https://github.com/pyca/cryptography}}} Python library, with ECDH and the ChaCha20\cite{bernstein2008chacha} stream cipher, respectivelly. Distribution of the DH public key is assumed outside of Fed-BioMed, offline, or trough a Public Key Infrastracture.

%% file: latex_files/evaluation.tex
In this Section we provide our theoretical and experimental evaluation of the two implemented SA protocols.

\textit{Complexity Analysis:}
JL's node computation is \(O(d)\), independent of the number of selected nodes, but requires modular exponentiation, and node communication for vector encoding is \(O(d \cdot 2 \cdot M)\) \cite{mansouri2022learning}. The server's computation is \(O(n + d)\), involving \(n\) multiplications and \(d\) exponentiation \cite{JL2013}.

LOM's node computation is \(O(nd)\), dependent on the number of selected nodes, using faster modular addition and PRF evaluation. The server's computation involves \(nd\) modular additions, and node communication is \(O(d \cdot M)\) \cite{kursawe2011privacy}.

\begin{figure}[h!]
 
  \begin{minipage}[b]{0.38\textwidth}
    \centering
    \label{tab:performance_metrics}
    \resizebox{\textwidth}{!}{
    \begin{tabular}{cccc}
    \toprule
    SA    & Time Train (s)   & Time Enc. (s)   & Time Tot. (s)     \\
    \midrule
    \multicolumn{4}{c}{FedIXI  $(d=246K; n_{\text{tot}} = n = 3)$}        \\
    JL    & $68.10 \pm 2.17$    & $52.21 \pm 0.85$   & $121.48 \pm 2.50$  \\
    LOM   & $46.51 \pm 1.39$    & $0.62 \pm 0.14$    & $48.22 \pm 1.03$   \\
    \midrule
    \multicolumn{4}{c}{FedHeart $(d=258; n_{\text{tot}} = n = 4)$}        \\
    JL    & $0.24 \pm 0.08$     & $0.08 \pm 0.01$    & $0.68 \pm 0.09$    \\
    LOM   & $0.20 \pm 0.09$     & $>0.01$            & $0.59 \pm 0.08$    \\
    \midrule
    \multicolumn{4}{c}{REPLACE-BG $(d=256K; n_{\text{tot}}= 180;n = 18)$} \\
    JL    & N/A                 & N/A                & N/A                \\
    LOM   & $53.72 \pm 8.61$    & $0.39 \pm 0.06$    & $57.42 \pm 6.95$   \\
    \midrule
    \multicolumn{4}{c}{FedProstate $(d=7.4M; n_{\text{tot}} = n = 4)$}    \\
    JL    & N/A                & $>300$               & $>300$               \\
    LOM   & $7.65 \pm 1.6$      & $9.22 \pm 0.38$    & $23.86 \pm 2.1$    \\
    \bottomrule
    \end{tabular}
    }
    
    \captionof{table}{}
    \label{table:results}
  \end{minipage}
  \begin{minipage}[b]{0.62\textwidth}

    \includegraphics[width=0.48\textwidth]{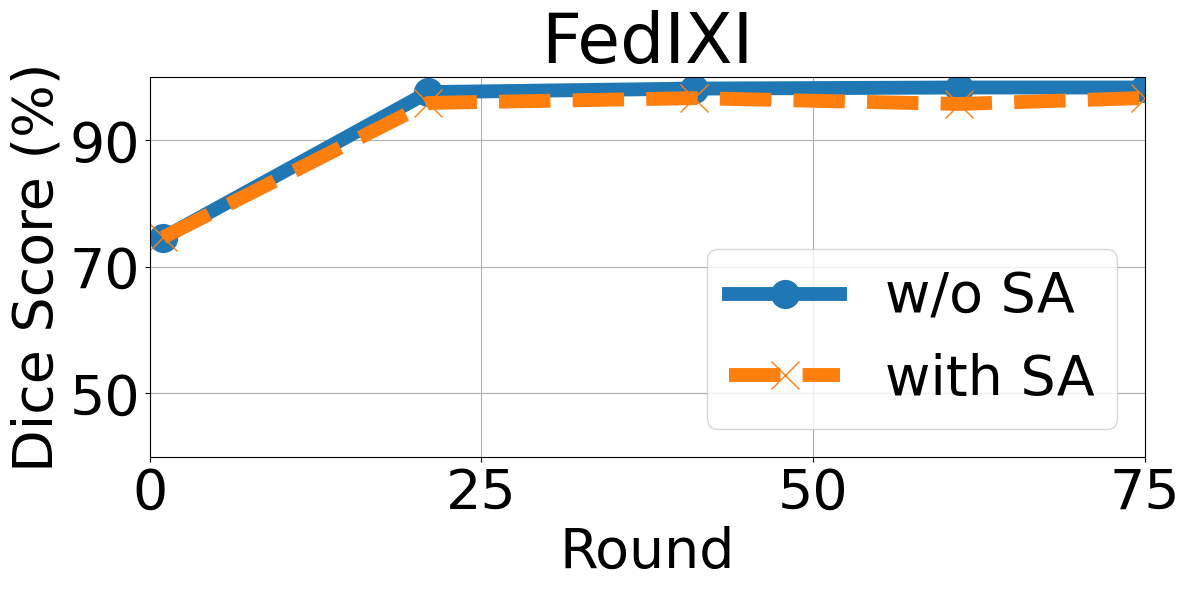} 
    \includegraphics[width=.48\textwidth]{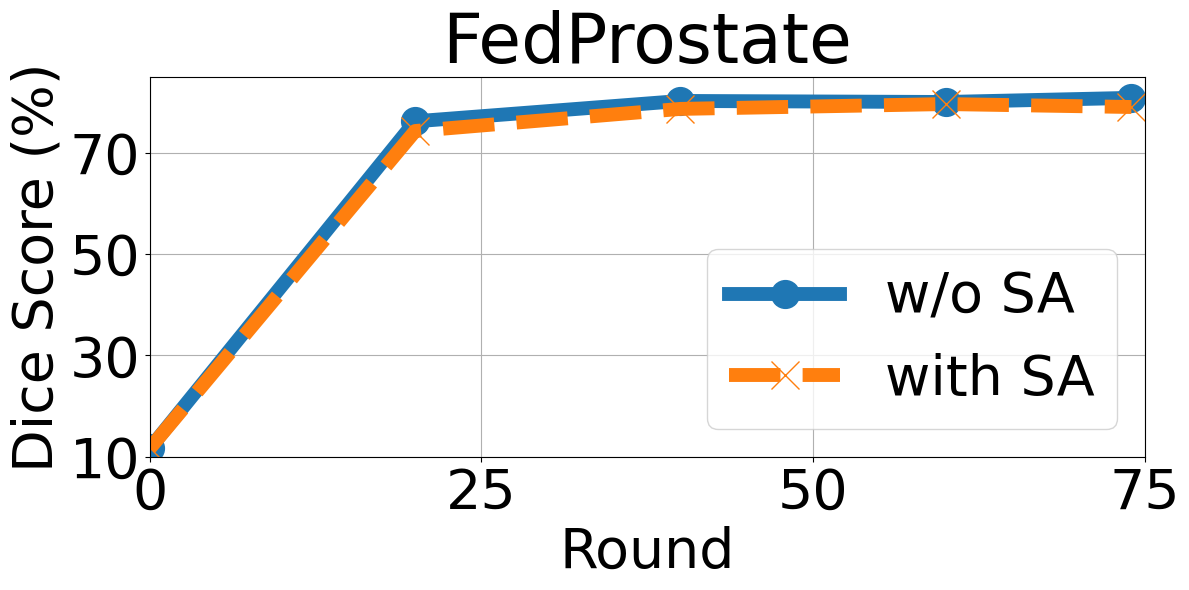} 
    \includegraphics[width=.48\textwidth]{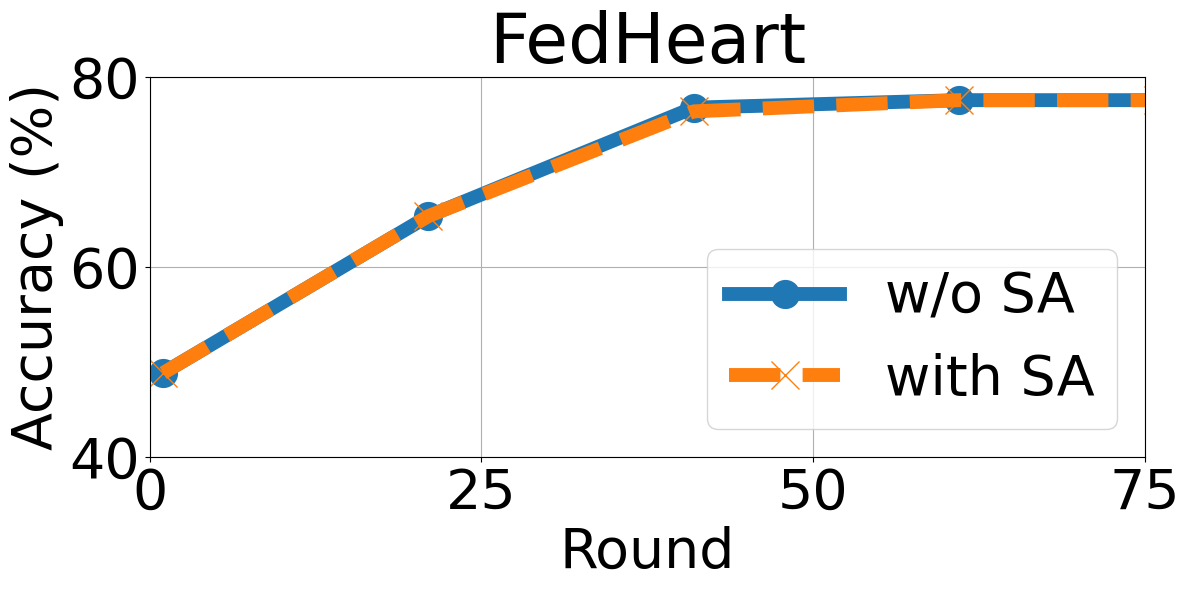}
    \hfill
    \includegraphics[width=.48\textwidth]{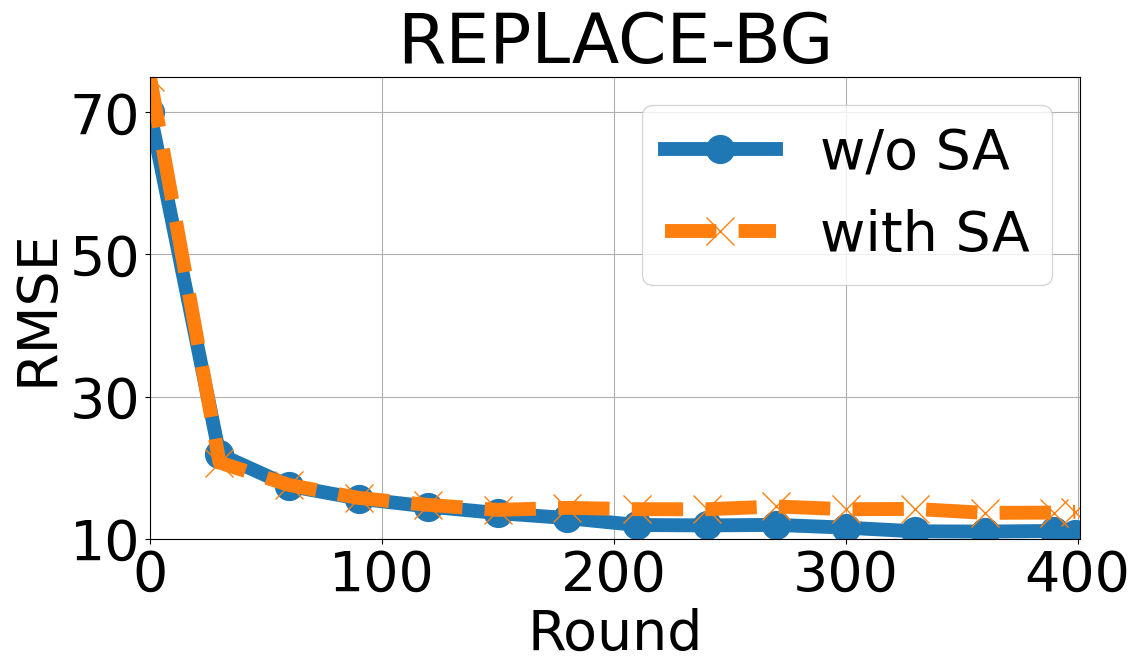}

                \captionof{figure}{}
    \label{fig:results}

  \end{minipage}
      {
        \renewcommand{\figurename}{}
            \captionsetup{labelformat=empty}

  \caption{
  Table \ref{table:results} Comparison of node average computation time across different SA protocols using four medical datasets. Each dataset is characterized by the total number of nodes ($n_{\text{tot}}$), the number of selected nodes ($n$), and the size of the local model ($d$).
  Fig. \ref{fig:results} Compare the task accuracy of the global model at each FedAvg aggregation with and without applying SA for FedIXI, FedProstate, FedHeart and REPLACE-BG. The SA is characterized by $L$ bits for representing the local model, $W$ bits for representing the maximum dataset size, and the specified maximum and minimum clipping range.}
  }
\end{figure}

\textit{Experimental evaluation:}
The experimental evaluation consists of tracking the computation time between  JL and the LOM.
We carried out the experiments by considering varying FL hyper-parameters represented by the number of total nodes $n_{tot}$, the number of selected nodes $n$, the number of FL rounds $T$, the number of local SGD steps $e$, the batch size $b$ and the learning rate $\eta$.
For SA, the hyper-parameterswe explored were the number of bits input $L$, the number of bits weight $W$. Moreover, we fixed the aggregation number of bits $M=32$ and the clipping range min and max.
Finally, we report the hardware  used to train ML model. We report all this information for each experiments in Table \ref{table:results}.

We use four medical datasets to evaluate the task accuracy of our SA implementations over the aggregated global model at each FL round, using a dedicated tasks-specific test set, and tracking the required computational resources for the nodes.

The four datasets are:

    $\bullet$ Fed-Heart \cite{ogier2022flamby}, providing patients' demography and clinical history from four hospitals. The task is to predict the clinical status of a patient (binary classification from tabular data). 
    For FL training and testing we follow \cite{ogier2022flamby}, and the target evaluation metric is the balanced accuracy.
    
    $\bullet$ Fed-IXI \cite{ogier2022flamby}, is composed by T1 and T2 brain magnetic resonance images (MRIs) from three hospitals. The task is supervised brain segmentation, and ground truth segmentations are provided. For FL training and testing we follow \cite{ogier2022flamby}, and the target evaluation metric is the dice score.
    
    $\bullet$ REPLACE-BG dataset \cite{aleppo2017replace} was obtained from a cohort of 202 participants. The task is prediction of  blood glucose levels for the subsequent hour based on data from the last three hours, including glucose levels, insulin boluses, and CHO content.

$\bullet$ FedProstate dataset \cite{innocenti2023benchmarking} provides T2 MRIs of the whole prostate from three publicly available datasets, and the task is supervised prostate segmentation. We defined the splitting criteria into different clients, the pre-processing methods, and the FL training and testing parameters coherently with \cite{innocenti2023benchmarking}.

Supplementary Table \ref{table:appendix} reports the dataset details, the FL and SA hyper-parameters, and the hardware specific for model training across experiments. The code is publicly available\footnote{\href{https://github.com/fedbiomed/fedbiomed}{\textcolor{blue}{GitHub code}}}.

In Table \ref{table:results}, we report the required node's computational resources comparing the two SA solutions. We present the average training time, encryption time, and total time. LOM consistently outperforms JL due to its faster underlying primitive (modular addition vs. modular exponentiation). Specifically, in all experiments where training runs on a CPU, LOM accounts for less than 1\% of the total time. When a GPU (e.g., in FedProstate) is available, the overall encryption time is around 40\% of the total time, considering a large input parameter dimension of $d=7.4M$.

In Figure \ref{fig:results}, we display the task accuracy comparison with and without SA for FedIXI, FedProstate, FedHeart and REPLACE-BG. This figure demonstrates that incorporating SA in Fed-BioMed affects the accuracy by no more than 2\% compared to the case without SA.


%% file: latex_files/appendix.tex
\newpage
\section{Appendix}
\input{protocols/protocol_v1}
\begin{table}[h!]
\centering
    \resizebox{0.8\textwidth}{!}{

\begin{tabular}{c@{\hskip 0.1in}c@{\hskip 0.1in}c@{\hskip 0.1in}c@{\hskip 0.1in}c@{\hskip 0.1in}c@{\hskip 0.1in}c@{\hskip 0.1in}c@{\hskip 0.1in}c@{\hskip 0.1in}c@{\hskip 0.1in}c}
\toprule
   Dataset         & \multicolumn{6}{c}{FL Hyper-params}                                                                                                & \multicolumn{3}{c}{SA Hyper-params}                                                                         &  Hardware spec.                            \\
   \cmidrule(lr){1-1}\cmidrule(lr){2-7} \cmidrule(lr){8-10}\cmidrule(lr){11-11}

     & $n_{tot}$ & $n$ & $T$ & $e$ & $b$ & $\eta$ & $L$ & $W$ & max/min &  \\
\midrule
FedIIXI     & 3      & 3  & 75                            & 10                            & 2                              & $1 \times 10^{-3}$            & 22                                    & 8                                     & +20/-20                     & CPU                          \\
FedHeart    & 4      & 4  & 75                            & 10                            & 8                              & $5 \times 10^{-4}$                             & 15                                    & 17                                    & +3/-3                       & CPU                          \\
REPLACE-BG  & 180    & 18 & 400                           & 10                            & 64                             &$1 \times 10^{-3}$                              & 13                                    & 15                                    & +3/-3                       & CPU                          \\
FedProstate & 4      & 4  & 75                            & 6                             & 8                              & $1 \times 10^{-3}$                              & 22                                    & 8                                     & +2/-2                       & GPU           \\
\bottomrule
\end{tabular}
}
\caption{FL hyper-params: number of total nodes $n_{tot}$, the number of selected nodes $n$, the number of FL rounds $T$, the number of local SGD steps $e$, the batch size $b$, and the learning rate $\eta$. SA hyper-params: number of bits input $L$, number of bits weight $W$ and clipping range max/min.}
\label{table:appendix}
\end{table}

%% file: protocols/protocol_v1.tex
\begin{figure*}
\centering
\raggedright
  \textbf{Prerequisites} Security paramter $\lambda$.
  
  \textbf{Parties:} Server, nodes ~$\users
$ and selected nodes $\selected$, s.t $|\users| = n_{tot}$ and $|\selected| = n$.
\raggedright
  
  \smallskip
  \scalebox{0.6}{
  \begin{minipage}[t]{.65\textwidth}
            \textbf{Public Parameters:} 
            \begin{itemize}
            \item  $(\perp, pp^{JL}) \gets \textbf{JL.Setup}(\lambda)$    
            \end{itemize}
                \textbf{Setup - Key Setup:}
                \begin{description}
                \item{Node $u$:}
                    \begin{enumerate}
                
                \item $sk_u \rgets \mathbb{Z}_{N^2}$. 
                \item $\{(v, [sk_u]_v)\}_{\forall v \in \users} \gets \tbf{SS.Share}(sk_u , t, \users)$
                \item Send $\forall v \in \users \setminus \{u\} , [sk_u]_v$ 
                \item Receive $\{[sk_v]_u\}_{\forall v \in \users \setminus \{u\}}$
                    \item  $[sk_0]_u \gets \sum_{\forall v \in \users}[sk_v]_u$
                \item Send $[sk_0]_u$ to Server
                    
                \end{enumerate}
                \item{Server}:
                \begin{enumerate}
                \item Collect $\{[sk_{0}]_u\}_{\forall u \in \users}$.
                    \item If $|\users| < t$, abort; otherwise, proceed.
                    \item $ sk_0 \gets \tbf{SS.Recon}(\{[sk_0]_v\}_{\forall v \in \users}, t)$
                \end{enumerate}

                \end{description}
  \textbf{Online - Protection ($\tau$):} 
\begin{description}
    \item{Node $u \in \users$:}
        \begin{enumerate}
            \item  $ \vencsync \gets \tbf{JL.Protect}(pp^{JL}, sk_u, \tau, \vinputsync)$
            \item Send $\vencsync$ to 
            
            \textit{Server}.
            \end{enumerate}
            \end{description}
            \textbf{Online - Aggregation ($\tau$):} 
            \begin{description}
                \item{Server:}
                \begin{enumerate}
                    \item Collect $\{\vencsync\}_{\forall u \in \users}$.
                    \item $\vaggsync \gets \tbf{JL.Agg}(pp, -sk_0, \tau, \{\vencsync\}_{\forall u \in \users}) $
                \end{enumerate}   
                \end{description}
                \vspace{1.1cm}

                                 \subcaption{JL}
                                                                             \label{fig:protocol_v1}

            \end{minipage}

\begin{minipage}[t]{.65\textwidth}
                \textbf{Public Parameters:} 
                \begin{itemize}
                \item $(\perp, pp^{LOM}) \gets \textbf{LOM.Setup}(\lambda)$ 
                \item $(\perp, pp^{KA}) \gets \textbf{KA.Param}(\lambda)$ 
                \end{itemize}
            \raggedright

              \textbf{Setup - Key Setup:}
                \begin{description}
                \item{Node $u \in \users$:}
                    \begin{enumerate}
                \item $(c_u^{SK},c_u^{PK}) \gets \tbf{KA.Gen}(pp^{KA})$
                \item Broadcast $c_u^{PK}$
                \item Receive $\forall v \in \users \setminus \{u\}, c_v^{PK}$
                \item  $\forall v \in \users \setminus\{u\}$,  $s_{u,v} \gets \tbf{KA.Agree}(pp^{KA}, c_u^{SK}, c_v^{PK})$             
            \end{enumerate}

                \end{description}
                \textbf{Online - Protection ($\tau$, $\selected$):} 
\begin{description}
    \item{Node $u \in \selected$:}
        \begin{enumerate}
            \item  $ \vencsync \gets \tbf{LOM.Protect}(pp^{LOM}, \{s_{u,v}\}_{\forall v \in \selected \setminus {u}}, \tau, \vinputsync)$
            \item Send $\vencsync$ to \textit{Server}.
            \end{enumerate}
                            \end{description}
                \textbf{Online - Aggregation ($\tau$):} 
                \begin{description}
                \item{Server:}
                \begin{enumerate}
                    \item Collect $\{\vencsync\}_{\forall u \in \selected}$.
                    \item $\vaggsync \gets \tbf{LOM.Agg}(pp^{LOM}, \{\vencsync\}_{\forall u \in \selected}) $
                \end{enumerate}   
                \end{description}

                	\subcaption{LOM}
                            \label{fig:protocol_v2}

\end{minipage}
            }
            
	\caption{SA protocols implemented in Fed-BioMed}
        \label{fig:sa_detailed}
    \end{figure*}